# Cosmology with ESO–SKAO Synergies

Mario G. Santos[1,2]
Stefano Camera[3,4,5,1]
Zhaoting Chen[6]
Steven Cunnington[7]
Jose Fonseca[8,1]

[1] Department of Physics & Astronomy, University of the Western Cape, Cape Town, South Africa
[2] South African Radio Astronomy Observatory, Cape Town, South Africa
[3] Department of Physics, University of Turin, Italy
[4] National Institute for Nuclear Physics, Turin, Italy
[5] INAF–Turin Astrophysical Observatory, Italy
[6] Institute for Astronomy, Royal Observatory, University of Edinburgh, UK
[7] Jodrell Bank Centre for Astrophysics, Department of Physics & Astronomy, University of Manchester, UK
[8] Institute of Astrophysics and Space Science, University of Porto, Portugal

We discuss the possible synergies for cosmology between SKAO and ESO facilities, focusing on the combinations SKA-Mid with the Multi-Object Spectrograph Telescope (4MOST) instrument built for ESO's Visible and Infrared Survey Telescope for Astronomy (VISTA) and SKA-Low with ESO's Extremely Large Telescope (ELT) multi-object spectrograph MOSAIC. Combining multiple tracers allows for tackling systematics and lifting parameter degeneracies. It will play a crucial role in the pursuit of precision cosmology.

## Introduction

The next decade will be dominated by large-sky surveys trying to improve on, and even provide a paradigm shift in, our understanding of the Universe. Such a change could manifest in clues about how to extend current theories of fundamental physics and combine the two pillars of quantum mechanics and general relativity. The standard model of cosmology — the so-called $\Lambda$-Cold Dark Matter model ($\Lambda$CDM) — with about six parameters, has been able to successfully describe almost all the measurements made so far. It assumes a universe with zero curvature, pressureless dark matter, and a cosmological constant, $\Lambda$, driving the observed accelerated expansion in recent times. In the recent past, observations using the cosmic microwave background (CMB) achieved unprecedented sub-percent errors on the parameters of this model (Planck Collaboration, 2020). This has been further confirmed through surveys of the large-scale cosmic structure, with the most recent examples being the Dark Energy Spectroscopic Instrument (DESI; DESI Collaboration, 2024) survey — a spectroscopic galaxy survey providing clustering statistics and constraints on baryon acoustic oscillations — and the Dark Energy Survey (DES; DES Collaboration, 2022) — a photometric galaxy survey capable of probing the matter distribution in the cosmos through weak gravitational lensing measurements.

There are, however, some tensions in the model (Di Valentino, Saridakis & Riess, 2022), in particular with the measured current expansion rate and the amplitude of cosmological fluctuations. Moreover, there is no theoretical reason to stick to the minimal $\Lambda$CDM model, and several extensions have been considered: a universe with non-zero curvature; dynamical dark energy models to account for the current accelerated expansion; interacting dark matter, including different scenarios for the neutrino species; modifications to general relativity; or changes in the nature of primordial fluctuations. Exploring possible inconsistencies in the standard cosmological model and constraining extensions to it is the main focus of upcoming cosmological surveys. Some examples are the Euclid satellite mission (Euclid Collaboration, 2024), or Vera C. Rubin Observatory (Izević et al., 2019). They will rely on a combination of probes, from galaxy clustering and weak lensing to Type Ia supernovae (SNIa). At the same time, the CMB is expected to continue to bring improvements and eventually detect primordial gravitational waves, providing insights into the physics of the early Universe.

As we enter the regime of high-precision cosmology, systematic errors are becoming the limiting factor in constraining cosmological parameters. Combining different experiments will help reduce such systematics and bring more confidence to any discovery while also lifting parameter degeneracies. Here we explore the synergy gains in combining data from the wide-area cosmology surveys planned with the SKAO and ESO facilities. For a description of what these surveys can do individually, please see SKA Cosmology Science Working Group (2020), and the 4MOST Cosmology Redshift Survey (CRS; Richard et al., 2019). Our focus is on clustering statistics using different tracers provided by these surveys. In particular, we shall consider the neutral hydrogen (HI) intensity mapping surveys planned for SKA-Mid/Low, the continuum galaxy surveys with SKA-Mid/Low, the 4MOST CRS and ELT-MOSAIC (Japelj et al., 2019).

## SKA-Mid HI intensity mapping and 4MOST CRS

Intensity mapping has been hailed as a highly efficient method to probe the large-scale cosmic structure (Santos et al., 2015). Instead of requiring the detection of single galaxies and the measurement of their redshift from their spectra, it looks for the total intensity from a given line, corresponding to the combined emission from all galaxies in a voxel. This is particularly useful for the HI line which is quite weak, allowing SKA-Mid to achieve high survey speeds in single-dish mode, while maintaining high spectral resolution. The MeerKAT telescope, a precursor to SKA-Mid, has already showcased the enormous potential of this technique (Wang et al., 2021, Cunnington et al., 2023, MeerKLASS Collaboration et al., 2024). The challenge is that excellent control of systematics is required, particularly those related to bright foreground emission. Cross-correlations with galaxy surveys can therefore be extremely useful in terms of dealing with these systematics, while at the same time improving the overall constraining power on cosmological parameters. Furthermore, using multiple tracers will improve constraints on primordial non-Gaussianity that cannot be obtained independently (Fonseca et al., 2015).

In the southern sky, one of the few wide spectroscopic surveys planned is the 4MOST CRS survey, in particular for $z < 1.0$, since at higher redshifts we expect Euclid to cover a large fraction of the southern sky. Therefore, we consider here a joint analysis between SKA-Mid HI intensity mapping and the 4MOST CRS luminous red galaxies (LRG) sample with a density of 400 galaxies per square degree. Given 4MOST's redshift coverage and area, we consider a large bin centred on $z = 0.55$ with width $\Delta z = 0.3$ and covering 7500 square degrees with complete overlap. For SKA-Mid, we assume a 20 000-square-degree survey, covering $0.3 < z < 3$ with 10 000 hours of total integration. Figure 1 shows the expected constraints on the spherical averaged power spectra and Figure 2 shows the resulting cosmological constraints. Although constraints from SKA alone are forecasted to be better, this assumes a control on systematics that might be too optimistic. A multi-tracer analysis on the overlapping area can provide similar constraints while being more robust in dealing with such systematics. This approach also helps to reduce cosmic variance on very large scales, achieving a 1-$\sigma$ error of 7 on the primordial non-Gaussianity parameter $f_{NL}$.

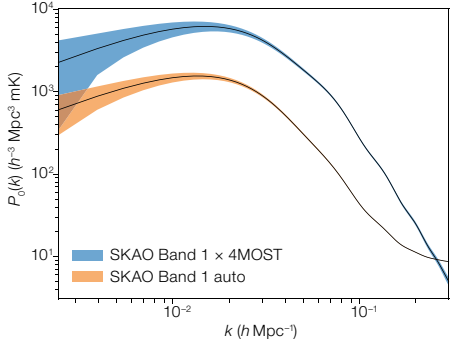

Figure 1. Constraints on the spherically averaged power spectrum for SKA-Mid and the cross-correlation with 4MOST.

### SKA-Mid continuum galaxy survey and 4MOST CRS

Catalogues of galaxies detected in the radio continuum typically extend to much higher redshifts than those detected at optical/near-infrared wavelengths, thanks to the fact that observations in the radio are not affected by dust obscuration. This translates into a long redshift tail, which can be exploited to track the evolution of critical cosmological quantities over cosmic time. However, redshift information is usually unavailable for continuum galaxies (see also Harrison, Lochner & Brown, 2017). For this reason, synergistic analyses with optical galaxy surveys have an enormous potential. For instance, by cross-identifying low-redshift radio-continuum galaxies with their optical counterparts, we can bin them tomographically, leaving the unmatched ones as an additional, mostly high-redshift, tomographic bin — an approach that has been suggested to deliver a large improvement on the Figure-of-merit (FoM) of the equation of state of dark energy (Camera et al., 2012).

Moreover, continuum galaxy surveys can cover up to three quarters of the sky, allowing one to probe huge swathes of cosmic volume. This is especially relevant in searches for primordial non-Gaussianity (Celoria & Matarrese, 2018) or for tests of general relativity on cosmological scales. In both cases, smoking-gun signatures appear only on the largest cosmic scales, making it paramount to be able to access the largest possible volumes. This can be achieved with an SKA-Mid continuum survey that is commensal with the planned intensity mapping survey by using the on-the-fly mapping technique. The synergy with area-overlapping spectroscopic galaxy surveys like 4MOST will be particularly important for going after such feeble signals (Raccanelli et al., 2012). Moreover, multi-wavelength data will unlock the possibility of performing a 'multi-tracer' analysis (Seljak, 2009), thanks to which the poor statistical sampling that plagues the largest scales can be alleviated. This approach has been investigated theoretically in the past in the context of continuum galaxy surveys with redshift/morphological cross-matching provided by optical/near-infrared spectroscopy. It has been suggested that it can provide one to two orders of magnitude better constraints on the amplitude of primordial non-Gaussianity, $f_{NL}$ (Ferramacho et al., 2014; Gomes et al., 2020).

Figure 2. Constraints on the angular-diameter distance, Hubble rate and the growth rate using the 4MOST CRS LRG sample and SKA-Mid Band-1 HI intensity mapping, and their multi-tracer analysis in the overlapping sky area. The quoted relative errors refer to the multi-tracer analysis. Only one large redshift bin was assumed, and all the bias parameters were marginalised over.

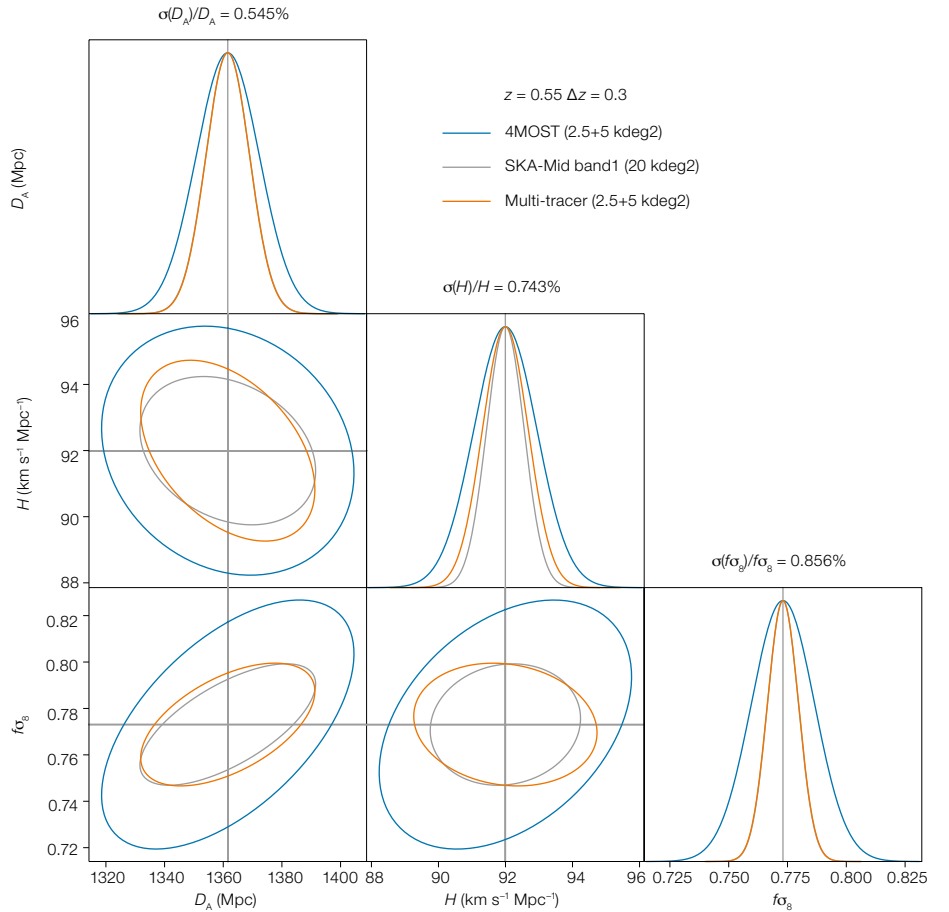

| | Frequency | Redshift | Survey Area | $\delta f$ | $A_{eff}/T_{sys}$ | Integration Time |
|---|---|---|---|---|---|---|
| SKA-Low | 272-320 MHz | 3.4-4.2 | 1 deg$^2$ | 250 kHz | 1.289 m$^2$/K | 1000 hrs |
| | <F> | NDIT | Total throughput | $N_{forest}$ | R | |
| ELT MOSAIC | 0.44 | 10 | 13% | 720 | 5000 | |

Table 1. The survey specifications of SKA-Low and ELT MOSAIC assumed in this work. From left to right, the top row lists the frequency range of the survey, the corresponding redshift, the effective survey area, the frequency resolution, the natural sensitivity of one station $A_{eff}/T_{sys}$, and the total integration time. The bottom row lists the mean transmitted fraction of the flux <F>, the number of integration times per field with each integration time lasting 1 h (NDIT), the total throughput, the number of forests observed, and the spectral resolution.

### SKA-Low HI intensity mapping and ELT-MOSAIC

The overlapping redshift range and survey area make the 21-cm signal and measurement of the Lyman-$\alpha$ forest ideal candidates for cross-correlations (Carucci, Villaescusa-Navarro & Viel, 2017). Cross-correlations are highly desirable to reduce the systematic effects in both probes and to produce tighter constraints on cosmological parameters, such as the growth rate, which can directly constrain theories of gravity. To forecast the detectability of the Lyman-$\alpha$ forest, we follow McQuinn & White (2011) to estimate the noise power spectrum. We assume the observations from the ELT MOSAIC instrument (Japelj et al., 2019). For the HI power spectrum, we simulate baseline distributions of SKA-Low for a tracking observation and calculate the corresponding thermal noise power spectrum according to Chen & Pourtsidou (2024). The specifications of the surveys are listed in Table 1. The redshift range of the cross-correlation is chosen to be 272–320 MHz, where observations of SKA-Low have low levels of radio frequency interference.

For the auto- and cross-power spectra, we adapt the modelling of Carucci, Villaescusa-Navarro & Viel (2017) with four fitting parameters: the HI bias $b_{HI}$, the HI redshift distortion parameter $\beta_{HI}$, the Lyman-$\alpha$ forest bias $b_F$, and the Lyman-$\alpha$ forest redshift distortion parameter $\beta_F$. The fiducials values are listed in Table 2.

To capture the anisotropy of the power spectrum in redshift space, it is desirable to split the k-space into clustering wedges of different $\mu = k_{\parallel}/k$, where $k_{\parallel}$ is the line-of-sight mode component. We choose the number of wedges $n_{wedge} = 5$ in this work. Furthermore, for the HI auto-power spectrum and the cross-power spectrum, we assume that the first wedge $\mu < 0.2$ is dominated by foregrounds and can not be used for parameter fitting. The resulting forecasts for the detectability of the power spectra are shown in Figure 3.

Using the forecasts of the power spectrum measurements, we perform MCMC parameter fitting to the model parameters. The 1-$\sigma$ confidence intervals of the model parameters are listed in Table 2. Instead of $\beta_{HI}$, we present the results in terms of the growth function $f = b_{HI} \beta_{HI}$, which is what we are interested in.

Including the cross-power spectrum helps to obtain better constraints on the HI parameters. Comparing the results to using auto-power only, they improve by ~ 30% the HI model parameters. These parameters are slightly biassed when cross-power is included. This is due to the fact that we assume a non-uniform cross-correlation coefficient and do not model it in the parameter fitting. Nevertheless, the fiducial values are within the 1-$\sigma$ confidence interval of the posterior. Cross-correlating HI and Lyman-$\alpha$ forests allow us to constrain the growth function $f = b_{HI} \beta_{HI}$ at $z \sim 4$, providing a novel probe of cosmological expansion at high redshifts.

### Summary

Near-term wide surveys in cosmology promise to bring the field to sub-percent accuracy and possibly unravel the nature of some of the most pressing open questions in cosmology, for example, the nature of dark matter and dark energy, and Inflation. In order to achieve this, we need to deal with complex systematics and break parameter degeneracies. Both issues can be addressed by combining different surveys with multiple probes. Focusing on SKAO and ESO instruments, we identified the combination SKA-Mid plus 4MOST CRS at $z < 1$ and SKA-Low plus ELT-MOSAIC at higher $z$ to be the most interesting for cosmology. Correlations with 4MOST will constrain the standard cosmological parameters through clustering statistics that should be more robust to systematics and help us characterise the nature of dark energy and primordial fluctuations. Correlations between SKA-Low and MOSAIC will improve our knowledge of the HI content of the Universe at $z \sim 4$ and provide a novel probe of cosmological expansion at high redshifts. Other possibilities not explored here are, for instance, the measurement of the redshift drift of objects following the cosmological expansion using the ArmazoNes high Dispersion Echelle Spectrograph (ANDES) at the ELT and an SKAO HI galaxy survey, albeit this would require much more collecting area than the current SKA-Mid setup (Rocha and Martins, 2023). Another interesting prospect would be the use of a single-dish telescope with spectroscopic capabilities in the submillimetre (like the proposed concept for the Atacama Large Aperture Submillimeter Telescope [AtLAST; Mroczkowski et al., 2024]). Combining CO and CII intensity mapping surveys from such a facility with SKA-Mid and Low will allow us to probe baryon acoustic oscillations and redshift-space distortions in the rather unexplored redshift regime between two and seven. One concerning conclusion from our analysis is the lack of spectroscopic coverage of the southern sky at $z < 1$. ESA's Euclid mission should be able to cover about 10 000 square degrees but mostly at $z > 1$ and Vera C. Rubin Observatory's Legacy Survey of Space and Time will only provide photometric redshifts. With the Extended Baryon Oscillation Spectroscopic Survey (eBOSS) and now DESI covering mostly the northern sky, the planned 4MOST CRS survey looks to be quite crucial for multi-wavelength cosmology in the southern hemisphere. In fact, it would be extremely powerful if such a survey could be expanded in the coming years in order to achieve number densities of at least 1000 per square degree in the redshift range $0.2 < z < 1.0$ over areas close to 10 000 square degrees. Although several competitive large-scale surveys will be coming online in the near future, it is becoming clear that combining these datasets will provide more than just the sum of its individual constraints and help us build robust measurements in our goal to narrow down the cosmological model.

| Parameter | $b_{HI}$ | $f = b_{HI} \beta_{HI}$ | $b_F$ | $\beta_F$ |
|---|---|---|---|---|
| Fiducial | 2.50 | 0.986 | −0.156 | 1.39 |
| Auto | $2.45^{+0.26}_{-0.27}$ | $1.053^{+0.504}_{-0.482}$ | $-0.156^{+0.002}_{-0.002}$ | $1.39^{+0.04}_{-0.04}$ |
| Auto+Cross | $2.51^{+0.19}_{-0.21}$ | $0.755^{+0.369}_{-0.354}$ | $-0.156^{+0.002}_{-0.002}$ | $1.39^{+0.04}_{-0.04}$ |

Table 2. The forecast constraints on the Lyman-$\alpha$ forest and 21-cm intensity mapping power spectra parameters using the assumed ELT-MOSAIC and SKA-Low surveys. The third row lists the results from using only the auto-power spectra signal while the last row includes cross-correlations.

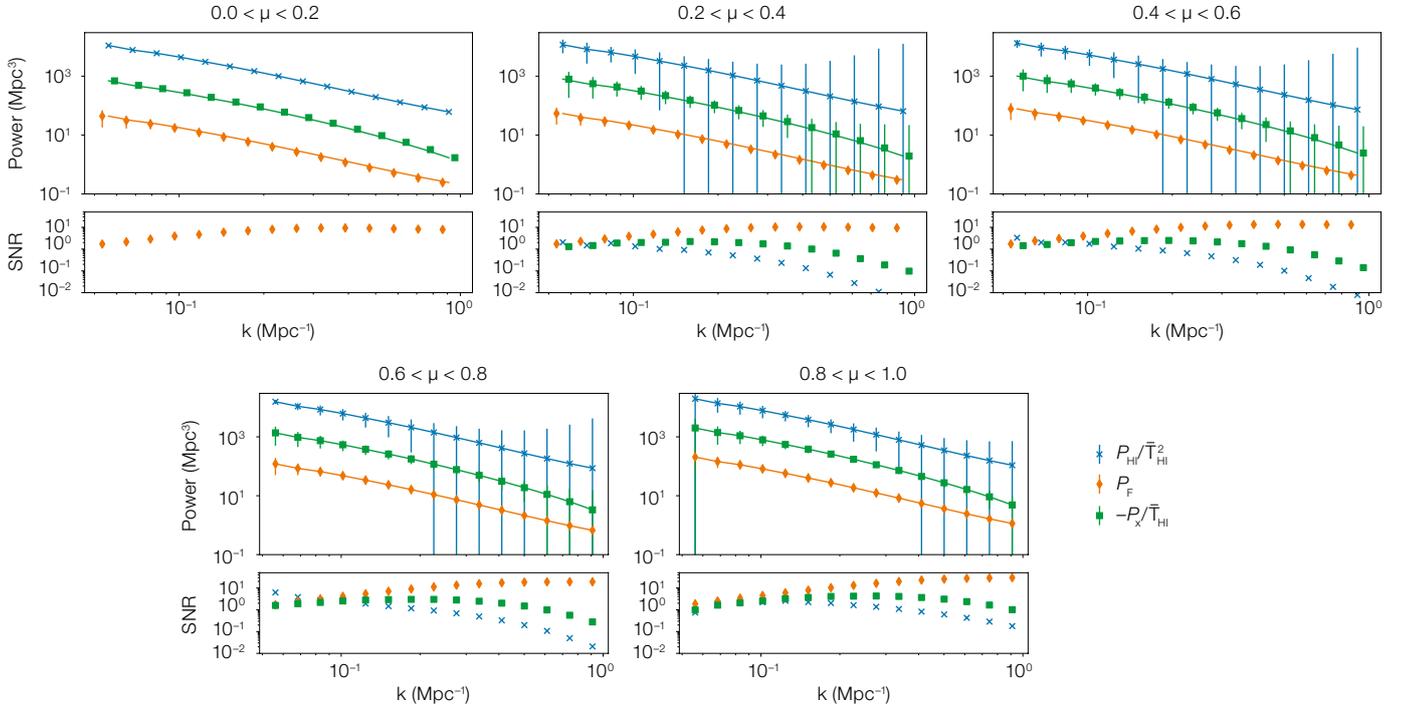

Figure 3. Forecasts for the 1D power spectrum measurements from Lyman-α forests using ELT MOSAIC and the 21-cm line using SKA-Low. The title of each panel shows the clustering wedge in which the power spectrum is averaged. The upper part of each panel shows the fiducial power spectrum for the 21-cm line, the Lyman-α forest and their cross-correlation. Note that the cross-power spectrum is negative, and its absolute values are shown. The centres of the k-bins for the Lyman-α forest and cross-power are misplaced by 5% for better visualisation. The error bar denotes the 1-σ of the measurement error. The lower part of each panel shows the signal-to-noise ratio of the measurements. We assume that the 21-cm power spectrum and the cross-power are dominated by foregrounds in the first wedge 0 < μ < 0.2.

Acknowledgements

MGS acknowledges support from the South African Radio Astronomy Observatory and National Research Foundation. SC acknowledges support from the Italian Ministry of University and Research (MUR), PRIN 2022 'EXSKALIBUR – Euclid-Cross-SKA: Likelihood Inference Building for Universe's Research', from the Italian Ministry of Foreign Affairs and International Cooperation (MAECI), Grant No. ZA23GR03, and from the European Union – Next Generation EU. JF thanks the support of Fundação para a Ciência e a Tecnologia (FCT) through the research grants UIDB/04434/2020 and UIDP/04434/2020 and through the Investigador FCT Contract No. 2020.02633.CEECIND/CP1631/CT0002. SCu is supported by a UK Research and Innovation Future Leaders Fellowship grant [MR/V026437/1].